# Accurate Depth-Resolved Temperature Profiling via Thermal-Radiation Spectroscopy: Numerical Methods vs Machine Learning

Dmitrii Shymkiv[1], Zhongyuan Wang[2], Brigham Thornock[1,3], Aiden Karpf[1,4], Camila Nunez[1,5], and Yuzhe Xiao[1*]

[1]*Department of Physics, University of North Texas, Denton, TX 76203, USA*
[2]*Department of Physics, The University of Texas at Austin, Austin, TX, 78712, USA*
[3]*Erik Jonsson School of Engineering and Computer Science, The University of Texas at Dallas, Richardson, TX, 75080, USA*
[4] *Physics and Astronomy Department, Pomona College, Claremont, CA 91711, USA*
[5]*Department of Computer Science, University of Virginia, Charlottesville, VA, 22904, USA*
*Email address: yuzhe.xiao@unt.edu*

**Abstract**

We present and compare three approaches for accurately retrieving depth-resolved temperature distributions within materials from their thermal-radiation spectra, based on: (1) a nonlinear equation solver implemented in commercial software, (2) a custom-built nonlinear equation solver, and (3) a deep neural network (DNN) model. These methods are first validated using synthetic datasets comprising randomly generated temperature profiles and corresponding noisy thermal-radiation spectra for three different structures: a fused-silica substrate, an indium antimonide substrate, and a thin-film gallium nitride layer on a sapphire substrate. We then assess the performance of each approach using experimental spectra collected from a fused-silica window heated on a temperature-controlled stage. Our results demonstrate that the DNN-based method consistently outperforms conventional numerical techniques on both synthetic and experimental data, providing a robust solution for accurate depth-resolved temperature profiling.

## I. INTRODUCTION

Thermal radiation is a fundamental physical phenomenon in which heated objects emit electromagnetic waves as a form of energy. Recent advances in thermal-radiation engineering have enabled a wide range of innovative applications [1], including radiative cooling [2], thermophotovoltaics [3], advanced infrared light sources [4], [5], and thermal camouflage [6]. Among the earliest and most widely adopted applications is infrared thermography, which utilizes infrared cameras to detect emitted radiation for non-contact temperature measurements. This technique is extensively used in medical imaging [7], defense [8], integrated circuit monitoring [9], and remote sensing [10]. However, conventional infrared thermography relies on the total radiated power to infer an effective surface temperature, limiting its ability to resolve subsurface temperature variations—an essential capability for applications such as thermal management in multilayer electronic devices [11] and three-dimensional (3D) medical diagnostics [7].

To address this limitation, we recently developed depth thermography, a non-invasive technique that enables temperature measurement as a function of depth [12]. Depth thermography utilizes spectrally resolved thermal radiation emitted from the semi-transparent spectral region of a material, where the emitted signal comprises contributions from various depths. By modeling the material as a multilayer structure with $N$ layers, the total radiated power can be expressed as:

$$I(\lambda) = \sum_{i=1}^{N} \bar{\epsilon}_i(\lambda) I_{BB}(\lambda, T_i), \tag{1}$$

where $\bar{\epsilon}_i(\lambda)$ is the local emissivity of the $i^{th}$ layer, $T_i$ is its temperature, and

$$I_{BB}(\lambda, T) = 2hc^2/\lambda^5 (e^{hc/\lambda k_B T} - 1) \tag{2}$$

is the blackbody spectral radiance according to Planck's law [12]. Here $h$ is Planck's constant, $c$ is the speed of light, and $k_B$ is Boltzmann's constant. Determining the temperature profile $T_i$ involves inverting Eq. (1)—an inherently ill-conditioned problem. This ill-conditioned nature significantly limits the accuracy and depth resolution of the reconstructed temperature distribution. In our previous work, we employed a brute-force search method and achieved a depth resolution of only 0.25 mm in a 1 mm-thick fused silica window [12]. Because experimental spectra inevitably contain noise, a robust inversion technique is essential to improve the accuracy and reliability of depth thermography.

An ill-conditioned problem is a mathematical or computational problem that does not meet the criteria of a well-conditioned problem, as defined by Jacques Hadamard [13]. According to Hadamard, a well- conditioned problem must satisfy three conditions: a solution must exist, be unique, and depend continuously on the input (i.e., be stable). In the context of temperature inversion from thermal-radiation spectra, even small noise in the measured spectra can lead to large deviations in the reconstructed temperature profile—an indication of solution instability. Furthermore, inverse problems are often inherently ambiguous [14], as multiple temperature distributions may produce spectral outputs that closely match noisy observational data [12]. Throughout the 20th century, numerous numerical and theoretical techniques were developed to mitigate the challenges of ill- conditioned problems, including methods such as regularization and minimization [14], [15], [16], [17], [18].

More recently, advances in machine learning — particularly deep learning and neural networks — have introduced powerful new approaches for solving inverse problems [19], [20], [21], [22]. These techniques have demonstrated significant potential across a wide range of disciplines, including optical metrology [23], spectroscopy [24], geophysics [25] and solid Earth geoscience [26]. In the context of temperature inversion, machine learning models have been successfully employed to retrieve temperature and humidity profiles from satellite observations [27] and ground-based microwave radiometer measurements [28] in remote sensing applications. Additionally, neural networks have been used to infer temperature and concentration distributions in combustion systems based on infrared emission spectra [29].

In this paper, we present and evaluate three methods for accurate temperature inversion from thermal-radiation spectra: (1) a nonlinear equation solver based on MATLAB's *lsqnonlin* function, (2) an inversion algorithm incorporating generalized cross-validation (GCV) [14], and (3) a deep neural network (DNN)-based approach. To assess the effectiveness of these methods, we first apply them to numerically generated datasets across three different structures. We then validate the approaches using experimental data obtained from a 1 mm-thick fused-silica window placed on a temperature-controlled heater. Our results show that the DNN-based method consistently delivers the highest accuracy, enabling robust depth-resolved temperature profiling through thermal-radiation spectroscopy.

## II. METHODS

Method 1 (hereafter referred to as MATLAB) utilizes *lsqnonlin*, a built-in MATLAB function designed for solving nonlinear least-squares problems [30]. This function minimizes the sum of squares of a set of nonlinear functions, making it suitable for curve fitting, parameter estimation, and complex optimization tasks. In the context of depth-resolved temperature profiling, *lsqnonlin* is used to find a set of temperatures ($T_i$), that minimizes the objective function

$$\sum_\lambda [Spectrum(\lambda) - \sum_{i=1}^{N} \bar{\epsilon}_i(\lambda) I_{BB}(\lambda, T_i)]^2 \,, \qquad (3)$$

where *Spectrum(λ)* is the input thermal-radiation spectrum provided to the solver. Also, *lsqnonlin* requires an initial guess as well as upper and lower bounds for temperature values. While this method can recover the exact temperature profile when the input spectrum is noise-free, its performance significantly deteriorates in the presence of noise, as demonstrated in Section III.

In method 1, the use of built-in MATLAB's *lsqnonlin* function limits our flexibility to customize the solver. Therefore, in method 2 (hereafter referred to as GCV), we developed a custom nonlinear equation solver based on regularization, an iterative approach originally formulated and tested for magnetotelluric inverse problems [14]. Following the notation in Ref. [14], the nonlinear ill-conditioned inverse problem can be expressed as

$$F(m) + \varepsilon = b, \tag{4}$$

where $b$ represents the noisy data (i.e., the measured spectrum), $\varepsilon$ denotes a measurement noise, $F$ is a nonlinear operator and $m$ is the solution of the inverse problem (i.e., the temperature distribution). The solution $m$ is obtained by minimizing the Tikhonov functional

$$\varphi = \|F(m) - b\|^2 + \beta \|W(m - m_{ref})\|^2, \tag{5}$$

where $m_{ref}$ is a reference solution and $\beta$ is the regularization parameter. This minimization is carried out iteratively using the damped Gauss–Newton method, starting from the initial guess $m = m_{ref}$. The regularization parameter $\beta$ is selected by minimizing the generalized cross validation (GCV) value:

$$GCV(\beta) = \frac{\|r(\beta) - r\|^2}{trace(I - C(\beta))}. \tag{6}$$

Further details of this algorithm can be found in Ref. [14]. For direct comparison with Method 1, we set $m_{ref}$ to be the same as the initial guess used in the MATLAB approach. Additionally, we impose constraints on the maximum and minimum allowable temperature values to mimic the upper and lower bounds applied in *lsqnonlin*.

For method 3 (hereafter referred to as DNN), we employ a conventional feedforward DNN model, with the input layer receiving the thermal-radiation spectrum, several hidden layers, and the output layer producing the temperature distribution. Unlike the MATLAB and GCV methods, which solve the inverse problem directly, the DNN must be trained to predict temperature profiles from given spectra. The training dataset is generated using Eq. (1), incorporating local emissivity and randomly generated temperature distributions within a certain range. The same local emissivity and temperature range are used in the MATLAB and GCV methods to calculate spectra and define upper and lower bounds of temperature.

To ensure a consistent comparison, all DNN models in this study share identical architectures and hyperparameters—except for the sizes of the input and output layers, which vary according to the problem. Each model consists of 20 hidden layers with 250 nodes per layer, employing the hyperbolic tangent (*tanh*) activation function. The datasets are split into training (60%), validation (20%), and test (20%) subsets, and normalized using mean normalization. Training is performed over 200 epochs using stochastic gradient descent, minimizing the mean squared error loss function.

### III. RESULTS
#### A. Numerical data
We first evaluate the performance of the three temperature-inversion methods using numerically generated data for three distinct structures: (1) a 1 mm thick fused-silica substrate on a heater stage,

(2) a 0.5 mm thick indium antimonide (InSb) substrate on a heater stage, and (3) a 5 μm thick gallium nitride (GaN) thin film on a sapphire substrate (Fig. 1). To demonstrate depth-resolved temperature inversion, the fused-silica structure is modeled as an 11-layer structure consisting of ten 0.1 mm thick silica layers atop a 200 nm thick aluminum layer. The InSb structure is modeled as a 21-layer system with twenty 0.025 mm thick InSb layers over a 200 nm aluminum layer. The GaN structure comprises 20 layers: ten 500 nm thick GaN layers, followed by five 1 μm thick sapphire layers and five 20 μm thick sapphire layers. We are mostly interested in the temperature distribution inside GaN as well as near the GaN/sapphire interface. Therefore, we intentionally introduce this non-uniform layer segmentation in the sapphire substrate. Depth thermography relies on thermal-radiation spectra in the semi-transparent region; accordingly, we focus on the wavelength ranges of 4.75–8 μm for fused silica, 5–10 μm for InSb, and 8–15 μm for GaN, with corresponding spectral resolutions of 50 nm, 10 nm, and 10 nm, respectively. The geometric and spectral parameters for these three test structures are summarized in Table I.

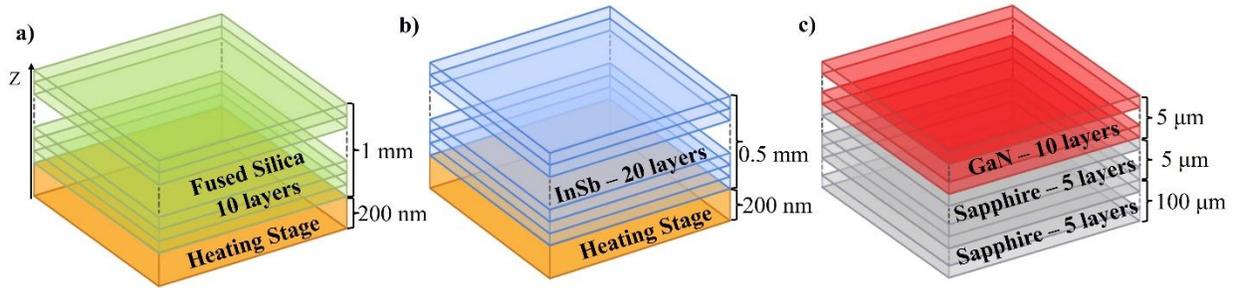

FIG 1: Three testing structures used to validate inversion methods via numerical data: a) 1 mm thick fused-silica substrate on a heater stage; b) 0.5 mm thick InSb substrate on a heater stage; c) 5 μm thick GaN film on a sapphire substrate.

Table I. Geometrical and spectral parameters of three testing structures.

|  | Fused silica | InSb | GaN |
|---|---|---|---|
| Material (bottom) | Aluminum | Aluminum | Sapphire |
| Number of layers | 1 | 1 | 5+5 |
| Total thickness | 200 nm | 200 nm | 5+100 μm |
| Material (top) | Fused Silica | InSb | GaN |
| Number of layers | 10 | 20 | 10 |
| Total thickness | 1 mm | 0.5 mm | 5 μm |
| Spectral range | 4.75–8 μm | 5–10 μm | 8–15 μm |
| Spectral resolution | 50 nm | 10 nm | 10 nm |

The optical properties (i.e., $n$ and $k$) of InSb, sapphire, and GaN are obtained from Ref. [31], [32], and [33], while those for fused silica are from our own ellipsometry measurement [12]. For each structure, the local emissivity of individual layers is calculated using the scattering matrix method and the dyadic Green's function [34]. Thermal-radiation spectra are then generated using Eq. (1), with layer temperatures $T_i$ randomly assigned within the range of 273 to 373 K. Some examples of temperature profiles and the corresponding thermal-radiation spectra for each structure are provided in Supplement Information, Sec. I. To simulate realistic experimental conditions, white Gaussian noise is added to the ideal spectra, with signal-to-noise ratios (SNR) set at 35, 45, and 50 dB to evaluate method performance under varying noise levels. For statistically meaningful analysis, each method is tested with $N$ = 100 different spectra and corresponding random temperature distributions. Method performance is assessed by the standard deviation (SD) for each layer across all test cases:

$$SD_j = \sqrt{\frac{1}{N}\sum_{i=1}^{N}(1 - T_{i,j}^{inv}/T_{i,j})^2} \times 100\%, \tag{7}$$

where $T_{i,j}^{inv}$ and $T_{i,j}$ represent the inverted and input temperatures for layer $j$ in the $i^{th}$ random temperature distribution.

Since the random temperature distributions are generated within the range of 273 to 373 K, these values are used as the lower and upper bounds for the numerical inversion methods (i.e., MATLAB and GCV). For DNN approach, 10,000 randomly generated temperature distributions within this range are used to train a model for each structure. For both the MATLAB and GCV method, a uniform temperature distribution equal to the top surface temperature is used as the initial guess and reference temperature, respectively. This choice ensures a fair comparison between the MATLAB and GCV methods. In practice, the surface temperature can be estimated by fitting the thermal-radiation spectrum in the opaque spectral region with a blackbody radiation curve [12].

The performance of the three temperature inversion methods is summarized in Fig. 2. As expected, the standard deviation increases with noise power across all structures and methods. One example of temperature inversion for fused-silica structure is shown in Supplement Information, Sec. II. For the fused-silica and InSb structures, the inversion error generally decreases with increasing layer number—that is, temperature estimates are less accurate for layers closer to the heater stage and more accurate for those near the top surface. This trend arises because thermal radiation from the deeper layers undergoes greater attenuation before escaping from the surface. In terms of radiative properties, layers near the bottom have smaller local emissivity values compared to top layers, resulting in smaller contributions to the total radiative power.

The GaN structure exhibits a more intriguing pattern of layer-dependent temperature inversion error (Fig. 2(f, i, l)). In addition to the near-exact retrieval of the top layer's temperature, two distinct dips in the error curves appear at layers 5 and 10. These dips, consistently observed across all inversion methods and noise levels, correspond to the interfaces within the non-uniformly segmented sapphire substrate. They arise from the relatively high local emissivity values of these layers. The dip at layer 10 results from both the contrast in optical properties between GaN and sapphire and the change in layer thickness—specifically, the GaN layer is 500 nm thick, whereas the adjacent sapphire layer is 1 μm thick. Moreover, the first five sapphire layers are significantly thicker than layers 6–10 (20 μm vs. 1 μm), which also contributes to their higher local emissivity and, consequently, lower inversion error. As a result, unlike the fused silica and InSb structures, the GaN structure does not exhibit a clear trend of decreasing inversion error with increasing layer number beyond the top five layers.

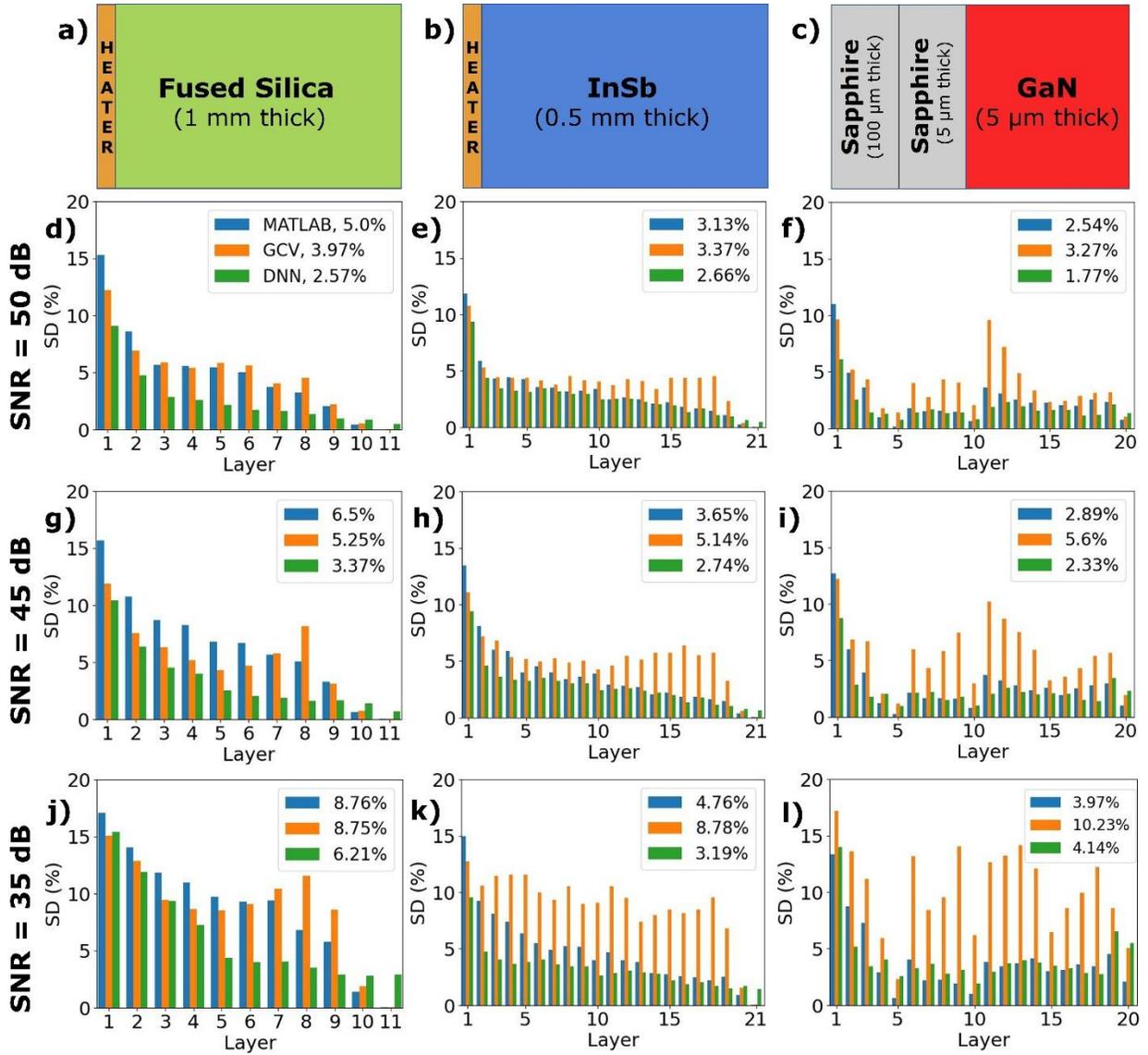

FIG 2: (a–c) Material distribution in three testing structures. Note that the layer thicknesses shown in these diagrams are not to scale. (d-l) Comparison of temperature inversion errors at different depths across three inversion methods with varying noise levels: (d-f) SNR = 50 dB; (g-i) SNR = 45 dB; (j-l) SNR = 35 dB. The legend indicates the layer-averaged standard deviation (SD).

For the fused-silica structure, the performance of the MATLAB and GCV methods is comparable across all noise levels, with standard deviations ranging from 4% to 9%. For the InSb and GaN structures, the MATLAB method yields lower average temperature inversion errors than the GCV method, particularly at SNR= 35 dB. Among all methods, the DNN models consistently achieve the best performance across all materials.

The performance of the DNN models can be further enhanced by incorporating noisy data and/or increasing the size of the training dataset. For example, in Fig. 3 we compare models trained with spectra containing different levels of noise. When the test set spectra exhibit noise levels similar to those used during training, prediction errors decrease relative to models trained on noise-free spectra. This improvement is especially evident under high noise conditions. Specifically, when tested on spectra with SNR = 35 dB, the model trained on SNR35 noisy data achieves more than twice the accuracy of the model trained on noise-free spectra. Furthermore, testing on noisy data with varying SNRs in Fig. 3 clearly shows that the layer-averaged error is minimized when the noise levels of the training and testing data match. These findings are consistent with previous studies on neural network performance under noisy and noise-free training conditions [19].

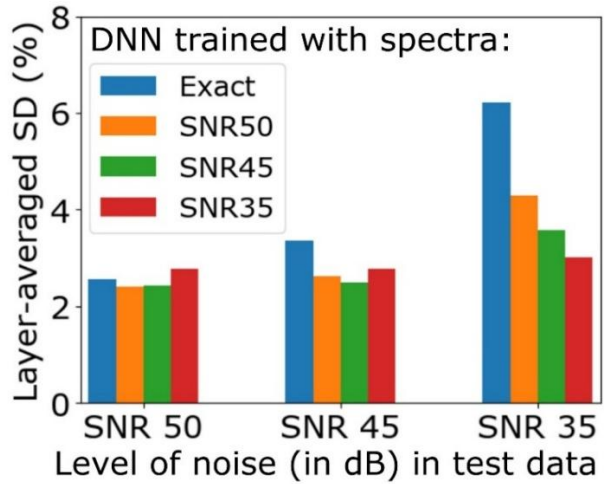

FIG 3. Comparison of layer-averaged temperature inversion errors for the fused-silica structure shown in Fig. 2(a) using DNN models trained with noisy and exact spectra.

### B. Experimental data

To further evaluate the performance of the three inversion methods, we apply them to experimental thermal-radiation spectra. These spectra were collected from a 1-mm-thick fused-silica window placed on a heater stage maintained at either 473 or 573 K, after steady-state thermal conditions had been established. A detailed description of the experiment is provided in Ref. [12]. When a bulk, isotropic, and homogeneous material reaches thermal steady state on a heated surface, its temperature is expected to increase linearly with depth. For the fused-silica window, we found the surface temperatures to be 468 and 556 K for heater stage temperatures of 473 and 573 K, respectively. These surface temperatures were measured using both an IR camera (since fused silica is opaque in the camera's operating spectral range of 7.5 to 13 μm) and spectral fitting within the opaque wavelength range, as described in Ref. [12].

The spectra were recorded with a resolution of 2 cm$^{-1}$, yielding 445 data points across the wavelength range of 4.75 to 8 μm. To further assess the performance of the inversion techniques, we investigate their accuracy at different depth resolutions — that is, how the fidelity of the reconstructed temperature profiles depends on the number of layers used in the model. All inversion methods are tested with three different layer counts: 6, 11, and 21, corresponding to depth resolutions of 0.2, 0.1, and 0.05 mm in the fused-silica substrate, respectively.

A uniform temperature distribution equal to the surface temperature is used as the initial guess for the MATLAB method and as the reference temperature for the GCV method. Upper and lower temperature bounds are set to ±50 K from the surface temperature to provide a reasonable estimated temperature range. For the DNN method, training datasets are generated using 10,000 random temperature profiles within the range of 506–606 K (or 418–518 K) for the 573 K (or 473 K) heater temperature case, across each depth resolution.

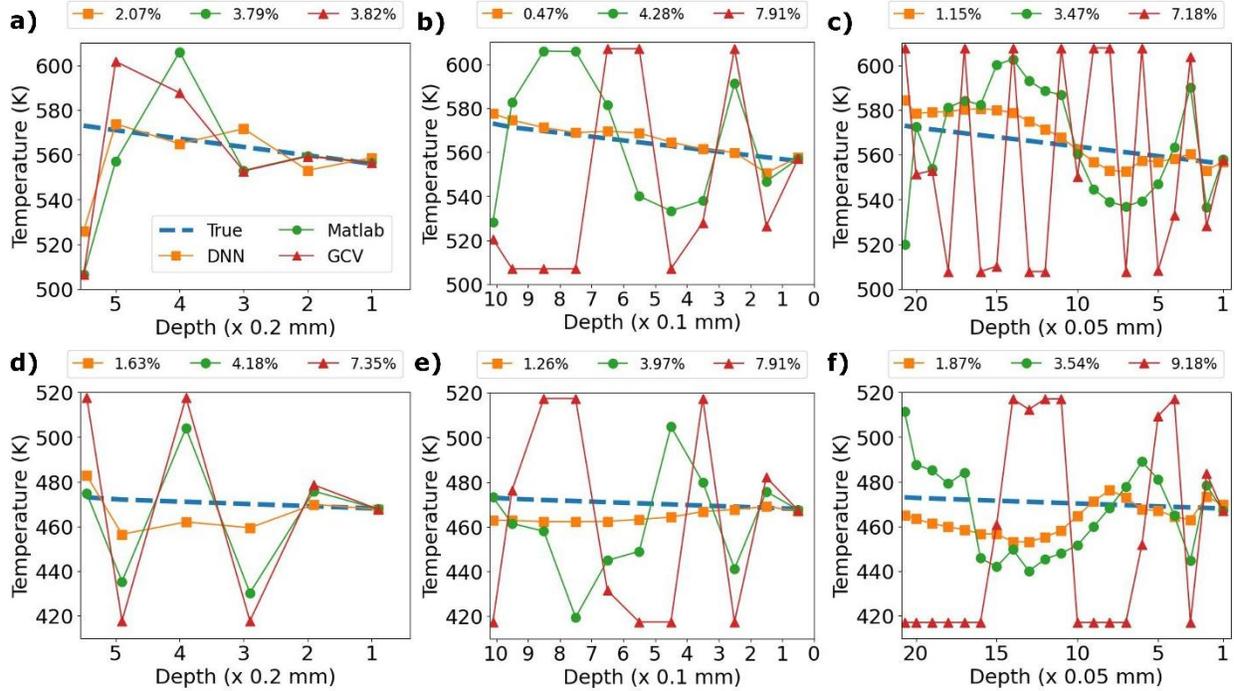

FIG 4. Comparison of inverted temperature profiles inside a 1 mm thick fused-silica window on a heater stage at either 573 K (a-c) or 473 K (d-f) across three methods with varying depth resolution: (a, d) 0.2 mm; (b, e) 0.1 mm; (c, f) 0.05 mm. The legend indicates the layer-averaged standard deviation (SD).

The depth-resolved temperature distributions retrieved from the experimental spectra of the fused-silica window are shown in Fig. 4. Among the three methods, the DNN approach yields the lowest temperature inversion errors across all tested scenarios. The MATLAB method outperforms the GCV approach. The errors associated with the MATLAB method remain consistent across all depth resolutions and both temperature settings, with an average error of about 3.8%. This is slightly better than the 5.0% error observed using numerical spectra (see Fig. 2(d)). We reference Fig. 2(d) here because the experimental spectra have a SNR close to 50 dB (see more discussions in Supplementary Information, Sec. III).

In contrast, the performance of the GCV method on the experimental spectra varies significantly with depth resolution. Its best performance occurs at a heater temperature of 573 K with a depth resolution of 0.2 mm, yielding an error of about 3.8%. However, in other cases, the inverted temperature profiles deviate substantially from the ground truth, with errors exceeding 7%. We attribute the subpar performance of the GCV method in this context to the limited dataset—only two experimental spectra were tested—which may lead to insufficient statistical representation.

The minimum temperature inversion error achieved is 0.47% using DNN model for the 573 K case with a depth resolution of 0.1 mm (Fig. 4(b)). Notably, this error is approximately five times lower than the 2.57% error obtained using numerical data (see Fig. 2(d)). Two main factors contribute to this discrepancy. First, the experimental spectra possess a significantly higher spectral resolution— approximately seven times greater—than that of the numerical spectra. As a result, the DNN model can predict the fixed-size output with improved accuracy. We believe this also partially leads to the improvement of the MATLAB method performance on experimental spectra versus numerical spectra. Second, the two error values are not directly comparable. The 2.57% error represents an average over 100 different test spectra, providing a statistically robust estimate, whereas the 0.47% error corresponds to a single instance of temperature inversion and lacks statistical generality.

It is evident that the DNN models with a depth resolution of 0.1 mm achieve the highest accuracy compared to models with resolutions of 0.2 and 0.05 mm across both temperature settings. While we do not claim that 0.1 mm is the optimal depth resolution for temperature inversion, the observed non-monotonic relationship between accuracy and resolution has computational origins. Specifically, the architecture and hyperparameters of all DNN models used for analyzing the experimental spectra were adopted from the model trained for numerical spectra with a 0.1 mm depth resolution. We suspect this contributes to the better performance observed at this resolution. Furthermore, from a numerical perspective, a model typically performs better when the output dimension is substantially smaller than the input. This consideration supports the expectation that the model would perform better at a 0.1 mm resolution than at a finer resolution of 0.05 mm.

As discussed in Sec. III.A (Fig. 3), the accuracy of the DNN models can be further enhanced by training it with noisy data. To explore this, we train the same DNN model—designed for a depth resolution of 0.1 mm and a heater temperature of 573 K—using synthetic spectra with varying SNRs. We then evaluate the performance of each trained model on the experimental spectrum. The results, presented in Fig. 5, show that the model trained with noisy spectra at an SNR of 50 dB performs best. Notably, our experimental spectra have an SNR of approximately 50 dB (see Supplementary Information, Sec. III). These findings are consistent with the results shown in Fig. 3 for the numerical spectra tested at SNR = 50 dB. Overall, we conclude that training the DNN with spectra having a noise level similar to that of the test data leads to improved inversion accuracy for both numerical and experimental datasets.

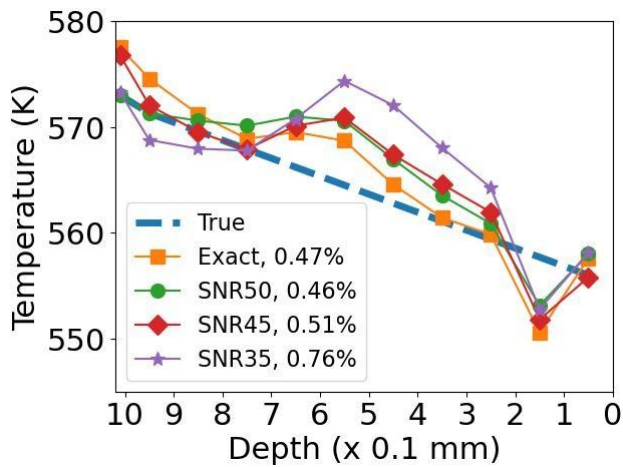

FIG 5. Temperature inversion results from the experimental spectrum of a fused-silica window placed on a heater stage at 573 K, using DNN models trained with exact spectra and noisy spectra at varying SNRs. The legend indicates the layer-averaged standard deviation (SD).

## IV. DISCUSSION AND CONCLUSION

Inverting depth-dependent temperature from thermal radiation is not a new concept. For instance, earlier studies have used multifrequency microwave radiometry to retrieve vertical profiles of temperature, water vapor, and cloud liquid water in the atmosphere [35], as well as internal temperature distributions in biological objects [36], [37], [38], [39]. In 1998, F. Solheim *et al.*

applied both neural network and conventional numerical methods for temperature inversion and found no significant advantage of the neural network approach over traditional techniques [35]. Their conclusion differs from our findings in this work. We attribute this discrepancy primarily to the limited computational resources available over two decades ago. Specifically, the 1998 study employed shallow neural networks with only one hidden layer, whereas our work utilizes a deep neural network architecture with 20 hidden layers.

Our results demonstrate that accurate depth-resolved temperature profiling is achievable across a wide range of materials — including both dielectrics and semiconductors—and over various depth scales (from millimeters to micrometers) and depth resolutions (from sub-millimeters to hundreds of nanometers). The key requirement is the presence of a spectral window in which the structure remains semi-transparent throughout its depth. Given the broadband nature of thermal radiation, such a window is usually easy to identify in practice. The depth resolution is not constrained by the diffraction limit and can be in the nanometer range. In practice, depth resolution is determined both by the number of spectral data points within the semi-transparent range and the total thickness of the structure. Even when this window is relatively narrow, we believe that, with sufficiently high spectral resolution, temperature inversion with high depth resolution across the entire structure remains achievable.

In this study, we use randomly generated temperature distributions to construct the training datasets. However, the training profiles can be customized to fit specific applications. For example, if the temperature is expected to vary monotonically with depth — either increasing or decreasing — the training set can be restricted to include only such profiles. While this limits the model's generality, it can enhance accuracy by incorporating additional physical constraints. This approach has been demonstrated in previous studies [29], [40], as well as in one of our recent work [41]. Additionally, to ensure a fair comparison, all DNN models in this study are implemented using identical architectures and hyperparameters (except for differences in input/output dimensions where necessary). Further fine-tuning of these parameters could potentially reduce prediction errors for specific material structures or particular combinations of spectral and depth resolutions.

In conclusion, we develop and compare three methods for inverting depth-resolved temperature distributions via thermal-radiation spectroscopy. Among them, the DNN approach consistently demonstrates the highest accuracy in temperature profiling, outperforming conventional numerical techniques including the GCV and MATLAB methods. In addition to its superior performance, the DNN approach offers greater flexibility—it can be further enhanced by training with noisy data, utilizing higher-resolution spectra (an option also available for the other two methods), or expanding the size of the training dataset. These advantages position the DNN method as the most promising approach for temperature inversion from thermal-radiation spectra.

**ACKNOWLEDGEMENTS**
D. S. and Y. X. acknowledge the support from DARPA YFA program (Grant No. D23AP0018900) and NSF LEAPS-MPS program (Award Number 2418002). B. T., C. N., and A. K. acknowledge the support from an NSF REU program hosted at UNT (Award Number 2244259).

# Supplementary Information

## I. Examples of numerical test data

Test data set for each of three considered structures in Section III.A consists of 100 random temperature profiles and corresponding thermal-radiation spectra. Noiseless spectra are calculated with Eq. (1) and white Gaussian noise with predefined signal-to-noise (SNR) ratio is added to generate noisy spectra with SNR = 50, 45 and 35 dB. Some examples of random temperature profiles and noiseless spectra for fused silica, GaN and InSb structures are shown on Fig. S1.

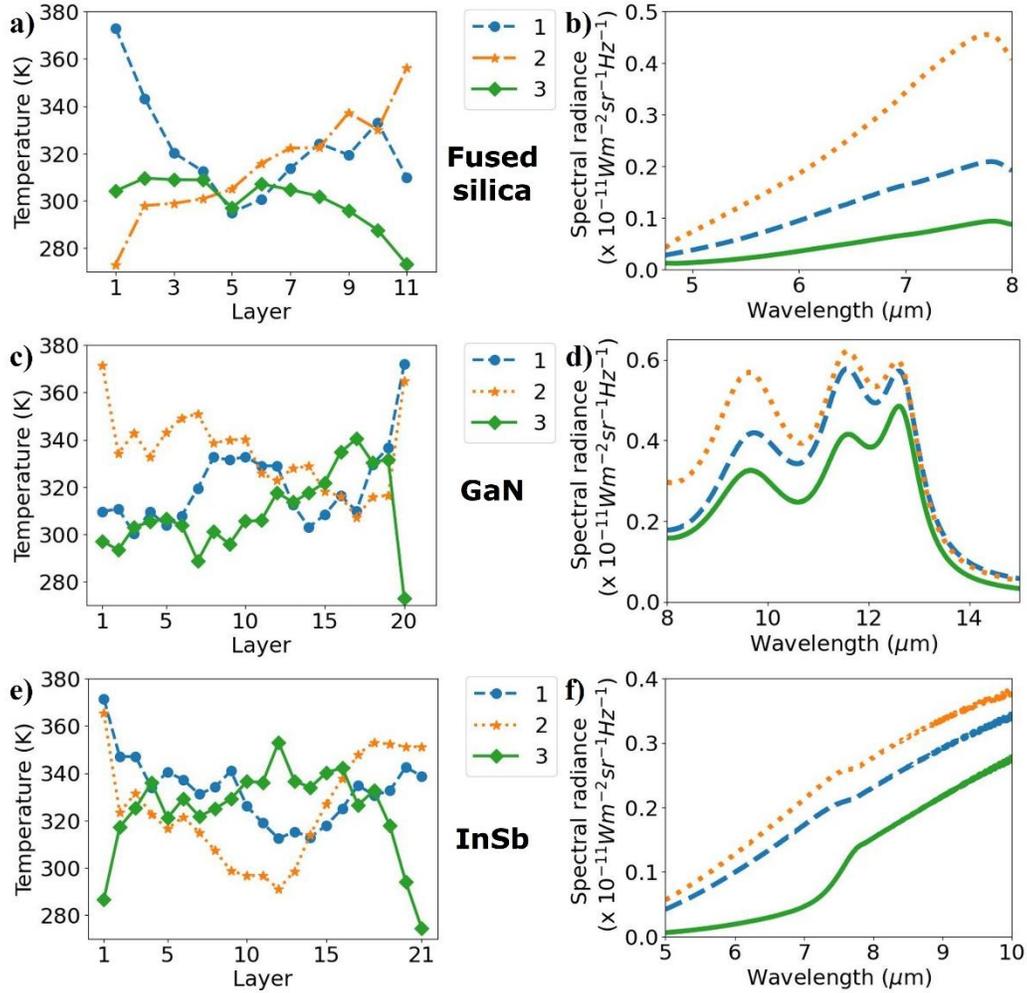

FIG S1. Examples of random temperature profiles (a, c, e) and corresponding thermal-radiation spectra (b, d, f) for three structures considered in Section III.A.

## II. Temperature inversion of fused-silica structure with numerical data

In this section, we show one example of the retrieved temperature profiles using three inversion methods for the fused-silica structure. The temperature profile and the corresponding thermal-radiation spectrum are shown in Fig. S1 (a-b), while the resulting inverted temperature profiles are plotted in Fig. S2.

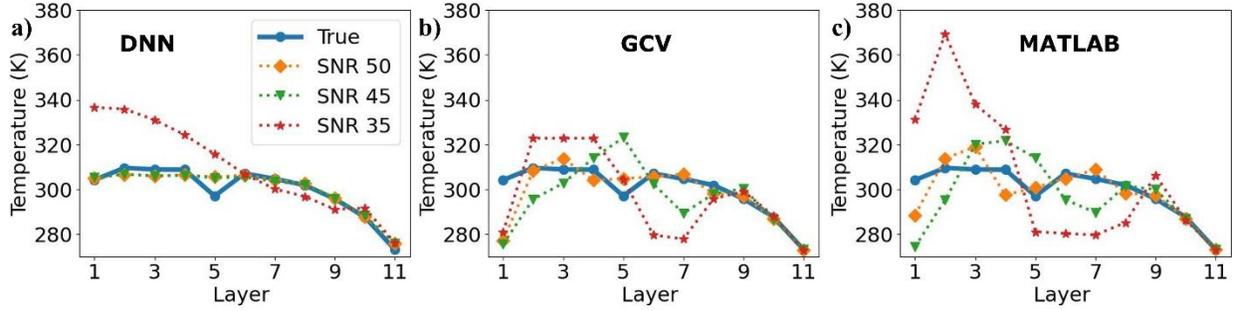

FIG S2. Results of three temperature inversion methods applied to the noisy thermal-radiation spectra corresponding to one temperature profile of the fused-silica structure (#3 in Fig. S1(a)).

### III: Noise level in experimental data

We estimated the SNR of the experimental spectra by analyzing the noise level in the measured spectra and comparing it to numerically generated spectra with varying levels of added Gaussian noise. Figure S3(a) shows a thermal-radiation spectrum measured from a blackbody reference (i.e., vertically aligned carbon nanotube (CNT) forest) placed on a heater at 373 K. At this temperature, the CNT sample exhibits negligible temperature gradients and is opaque across all wavelengths, allowing its spectrum to be well-fitted by blackbody radiation curve at 372.3 K.

To quantify the SNR in dB, we selected two spectral regions corresponding to the edge and peak of the spectrum (left and right sides of Fig. S3(a)). The measured wavelength-averaged noise level is 0.57% near 5 μm and 0.08% near 14 μm. Averaging these values yields a noise level of 0.325%, consistent with an SNR of 50 dB, which corresponds to a noise amplitude of about 0.3%.

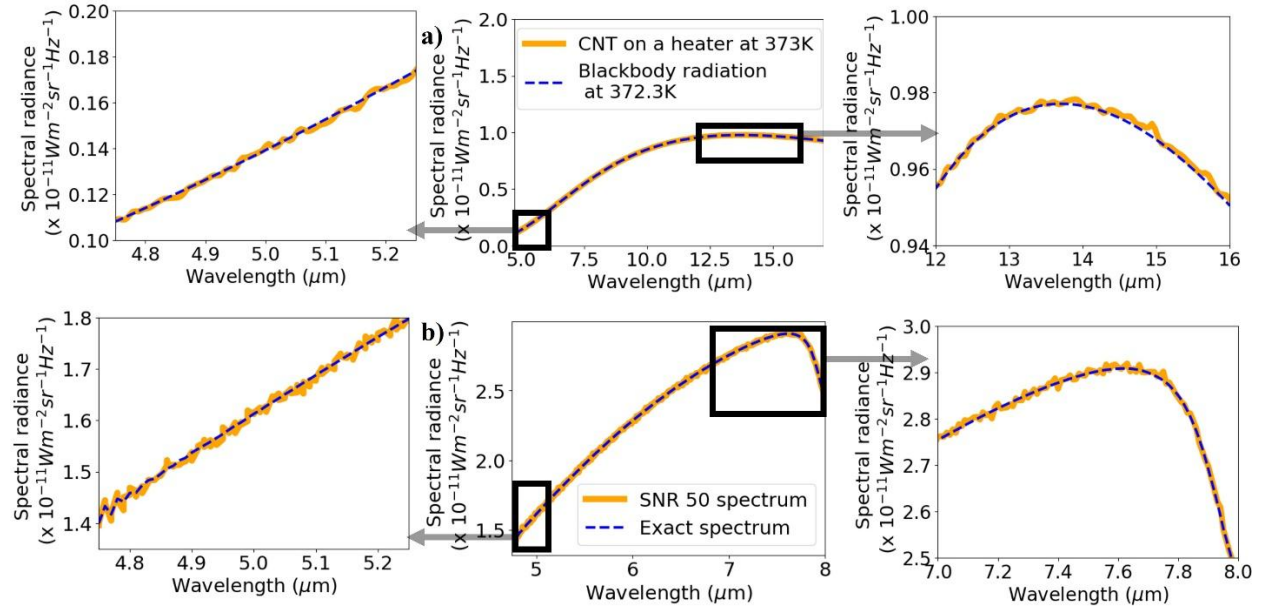

FIG S3. (a) Experimental thermal-radiation spectrum of a CNT laboratory blackbody placed on a heater at 373 K (solid orange) and fitted blackbody radiation spectrum at 372.3 K (dashed blue). (b) Exact numerical thermal-radiation spectrum (dashed blue) and noise-added spectrum with SNR = 50 dB (solid orange). Zoomed-in spectra near the edge and peak are shown on the left and right, respectively.

To further confirm this, we also analyzed the noise level distribution using a numerically generated spectrum. Figure S3(b) shows an arbitrarily generated spectrum with added white Gaussian noise at an SNR of 50 dB. The wavelength-averaged noise level near the edge is 0.41%, while near the peak it is 0.18%. The average of these values, around 0.3%, is consistent with the expected noise level for an SNR of 50 dB.